%
\input epsf
\font\tita=cmr10 scaled \magstep 2
\overfullrule=0pt
\baselineskip 18pt plus 2pt minus 2pt
\hsize=6.23truein
\vsize=8.5truein
\parindent=20pt 
\centerline {\tita Measurement of negative particle multiplicity}
\medskip
\centerline {\tita in S-Pb collisions at 200 GeV/c per nucleon}
\medskip 
\centerline {\tita with the NA36 TPC}
\vskip 1.5truecm
\centerline{\tita NA36 Collaboration}
\vskip 1truecm
\noindent
E. Andersen$^{1,a}$, R. Blaes$^2$, J.M. Brom$^2$, M. Cherney$^3$, 
B. de la Cruz$^4$, C. Fern\'andez$^{5,}$\footnote{*}{Deceased.}, 
C. Garabatos$^{5,b}$, J.A. Garz\'on$^5$, W.M. Geist$^2$, D.E. Greiner$^6$, 
C.R. Gruhn$^{6,c}$, M. Hafidouni$^{2,d}$, J. Hrubec$^7$, P.G. Jones$^6$, 
E.G. Judd$^{8,e}$, J.P.M. Kuipers$^{9,f}$, M. Ladrem$^{2,g}$, 
P. Ladr\'on  de Guevara$^4$, G. L{\o}vh{\o}iden$^1$, J. MacNaughton$^7$, 
J. Mosquera$^5$, Z. Natkaniec$^{10}$, J.M. Nelson$^8$, G. Neuhofer$^7$, 
C. P\'erez de los Heros$^{4,h}$, M. Pl\'o$^5$, P. Porth$^7$, B. Powell$^9$, 
A. Ramil$^5$, H. Rohringer$^7$, I. Sakrejda$^6$, 
T.F. Thorsteinsen$^{1,\!\displaystyle *}$, 
J. Traxler$^7$, C. Voltolini$^2$, K. Wozniak$^{10}$, A. Ya\~nez$^5$ and 
R. Zybert$^{8,i}$.
\vskip 1truecm
\item{1)} University of Bergen, Dept. of Physics, N-5007 Bergen, Norway.
\item{2)} Institut de Recherches Subatomiques (IReS), IN2P3-CNRS/Universit\'e 
Louis Pasteur, B.P. 28, F-67037 Strasbourg Cedex 2, France.
\item{3)} Creighton University, Dept. of Physics, Omaha, Nebraska 68178, 
USA.
\item{4)} CIEMAT, Div. de F\'{\i}sica de Part\'{\i}culas, E-28040 Madrid,
Spain. 
\item{5)} Universidad de Santiago, Dpto F\'{\i}sica de Part\'{\i}culas, 
E-15706 Santiago de Compostela, Spain.
\item{6)} Lawrence Berkeley Laboratory (LBL), Berkeley CA 94720, USA.
\item{7)} Institut f\"ur Hochenergiephysik (HEPHY), A-1050 Wien, 
Austria.
\item{8)} University of Birmingham, School of Physics and Space Research, 
Birmingham B15 2TT, UK.
\item{9)} European Organization for Nuclear Research (CERN), CH-1211 
Gen\`eve 23, Switzerland.
\item{10)} Instytut Fizyki Jadrowej, PL-30 005 Krak\'ow 30, Poland.
\item{a)} Present address: Haukeland Sykehus,  Med.-tek. avd., N-5021 Bergen, 
Norway.
\item{b)} Present address: GSI, Darmstadt, Germany.
\item{c)} Present address: CERN, PPE division.
\item{d)} Present address: 50, Rue de Soultz, 67100 Strasbourg, France.
\item{e)} Present address: LBL, USA.
\item{f)} Present address: ETH, Z\"urich, Switzerland.
\item{g)} Present address: Ecole Normale Sup\'erieure, Algiers, Algeria.
\item{h)} Present address: University of Uppsala, Uppsala, Sweden.
\item{i)} Present address: ZYBERT Computing Ltd, Birmingham B32 3PW, UK.
\vfill\eject
\noindent
{\bf Abstract.}\quad A high statistics study of the negative particle 
multiplicity distribution from S-Pb collisions at 200 GeV/c per nucleon 
is presented. The NA36 TPC was used to detect
charged particles; corrections are based upon the maximum entropy method.
\vskip 2truecm 
\noindent
{\bf Introduction.}
\bigskip
Experiments with ultrarelativistic nucleus-nucleus collisions may lead to
the discovery of a new state of matter: the quark gluon plasma. Various 
experiments involved in this program are both investigating different 
predicted plasma signatures and trying to understand the dynamics of these 
collisions in general. With this in mind various features are studied; 
among them multiplicity distributions. 

Here a measurement of the inclusive multiplicity distribution of negative 
particles produced in S-Pb collisions at 200 $GeV/c$ per nucleon [1] is 
presented; it was obtained by the NA36 Time Projection Chamber (TPC) 
designed for the high multiplicity environment of heavy-ion collisions. 
Detection inefficiencies were corrected for using a maximum entropy method.  
\bigskip\noindent
{\bf The experimental apparatus.}
\bigskip
The NA36 spectrometer consisted of a TPC [2] and a set of detectors originating
from the former European Hybrid Spectrometer [3]. Only the detectors relevant 
to the multiplicity measurement are described here.
The Pb target (5\% of an interaction length) is surrounded by two triggering 
Si counters located 20 $cm$ upstream and downstream. The one upstream selects 
the incident beam S ions, whereas the one downstream rejects those S ions 
having not interacted in the target. This rejection, based on the amplitude 
of the signal, defines the "minimum bias" event sample.
The target and the Si detectors are placed inside a vacuum tank extending 
up to the TPC in order to minimize secondary interactions. 

The TPC has a volume of $50 \!\times\! 50 \!\times\! 100 \,\,cm^3$; its 
electric field is parallel to a strong horizontal magnetic field (2.7 $T$) 
produced by a superconducting magnet. The TPC is installed 1 $cm$ above 
the beam line and 1 $m$ downstream of the target in order to avoid most
nuclear fragments in its sensitive volume. Because of the strong magnetic 
field and the position of the TPC, most of the detected particles have 
the same charge sign.
The TPC is read out by a matrix of 1 $cm$ long anode wires, organized in 40
vertical rows of 192 wires each. The wire and row pitches are 0.1 $inch$ and 
1 $inch$, respectively. Two-track resolutions of $\sigma_Y^t=5 \,\,mm$ and 
$\sigma_Z^t=10 \,\,mm$, in the vertical direction of the bend plane and the 
drift direction, respectively, were obtained [2]. Data were taken with two 
magnetic field polarities.

The acceptances of the TPC in terms of laboratory rapidity $y$ and transverse 
momentum $P_T$ are shown in the figure 1. They were derived from Monte-Carlo 
simulations to be discussed below. In the following only the vertex tracks   
with $2 \leq y \leq 5$, $0.1 \leq P_T \leq 2 \,\,GeV/c$ and  
$-89^\circ \leq \varphi \leq 89^\circ$ ($\varphi$ is the azimuthal angle) 
will be considered. These restrictions were made in order to avoid large 
corrections. 
\bigskip\noindent
{\bf Selection of events.}
\bigskip
The results presented here are based on about 14000 minimum bias events.  
The hits recorded by the TPC were used to reconstruct and fit tracks 
by the usual track road and track following methods [4].
The position of the primary vertex was determined by the Kalman filter 
method [5]. Its distribution along the beam axis (Fig. 2)
agrees with Monte-Carlo simulations. Contamination from interactions outside 
the target has been removed from our data sample by rejecting vertices 
with $X < -115 \, cm$ and $X > -109 \, cm$.

Contamination from electromagnetic dissociation (EMD) was removed from the  
sample of events without any vertex tracks 
observed in the TPC. Its contribution was calculated from the charge changing 
and the production cross-sections measured by NA36 [6] and estimated 
elsewhere [7]; about $36.5 \pm 6.1\%$ of the minimum bias events are due to 
EMD. All results reported in the following refer therefore, to the strong 
inelastic cross section.
\bigskip\noindent
{\bf Correction of the multiplicity distribution.}
\bigskip
The observed negative multiplicity distribution has to be corrected for 
efficiency and geometrical acceptance of the TPC, as well as for inelastic
reinteractions in the target and downstream of it.
For an observed multiplicity
distribution $O$ of $M$ bins and the true one, $T$, of $N$ bins one arrives 
at the following relation [8,9]:
$$O_m = \sum_{n=1}^{N} P_{mn} T_n, \eqno(1)$$  
where $O_m$ ($m = 1,~..., M$) is the fraction of events with observed 
negative multiplicity in the bin $m$ and $T_n$ is the fraction of events 
with true negative multiplicity in the bin $n$; 
$P_{mn}$ is the probability that an event with a true multiplicity in the bin 
$n$ be observed as an event with observed multiplicity in the bin $m$. 
Hence, the $P_{mn}$ satisfy the following relation:   
$$\sum_{m=1}^{M} P_{mn} = 1.                                     \eqno(2)$$
\smallskip
The matrix $P$ was determined from a Monte-Carlo simulation of S-Pb 
interactions, based on 23000 IRIS events [10], and which included
a GEANT simulation of reinteractions in all parts of the experimental setup
(target, TPC, ...), as well as a simulation of signal formation in the TPC 
and of   
reconstruction and analysis programs. The momenta of simulated 
tracks at the primary vertex were used to establish the acceptances discussed 
above.

The underconstrained system of equations (1) cannot be solved in a 
straightforward way; this may lead to some negative values of $T_n$. Since 
the bins' contents are subject to statistical fluctuations, one should, 
rather, look for a solution ($T_1, ..., T_N$) that describes the data well 
in a statistical sense, which means that the differences 
$\left\vert O_m - \sum_{n=1}^N P_{mn} T_n \right\vert$ are of the order of 
the corresponding statistical errors $\sigma_m$ and not equal to zero as 
would be the case if (1) is solved directly. 

The usual procedure (least squares fit) consists in minimizing
$$\chi^2=\sum_{m=1}^M\left({{O_m-\sum_{n=1}^N P_{mn} T_n}\over{\sigma_m}}\right)^2.\eqno(3)$$  
This method yields unstable solutions [8,9,14].

An alternative method to choose one probability distribution from a set of 
distributions compatible with the data requires maximizing the Shannon 
entropy [11]
$$S=-\sum_{n=1}^N t_n\,\ln t_n,\quad\quad t_n = {{T_n}\over{\sum_{i=1}^{N} T_i}} \eqno(4)$$ 
under constraints imposed by the data. This is known as the principle of 
maximum entropy, proposed by E.T. Jaynes [12]. It has been shown that the  
maximum entropy method (MEM) is the only consistent method of inference for 
underconstrained problems [13]. The MEM has been used earlier for the 
correction of measured multiplicity distributions [8,9,14,15]. 

Here, the observed multiplicity was corrected as in ref. [14]. The constraints 
were chosen by requiring that the moments of the observed distribution be 
reproduced by the true one, i.e.
$$\sum_{m=1}^M m^q O_m = \sum_{m=1}^M \sum_{n=1}^N m^q P_{mn} T_n  \eqno(5)$$
for some values of $q$. The choice of moments was motivated by the fact that
they are less sensitive to statistical fluctuations than individual bins. 
The number of constraints (i.e. number of different moments) was selected such  
that 
$$\chi^2 \approx M,                                               \eqno(6)$$ 
where the $\chi^2$ is given by (3). A set of seven constraints corresponding to 
$q = -1, 0, 1, ..., 5$ was necessary.  
\bigskip\noindent
{\bf Discussion of the results.}
\bigskip
Figure 3 shows the multiplicity distribution as it is obtained after the  
removal of the EMD contamination. This distribution will be corrected in the 
following in the TPC phase space: $2 \leq y \leq 5$, 
$0.1 \leq P_T \leq 2 \,\,GeV/c$, $-89^\circ \leq \varphi \leq 89^\circ$, and 
in the full phase space. For each case the appropriate matrix $P_{mn}$ was 
determined.

Figure 4 shows the corrected multiplicity distribution in the
TPC phase space.
The error bars are statistical. The mean negative multiplicity and the
dispersion $D_{-}$
$=\sqrt{<\!\!n_{-}^2\!\!> - <\!\!n_{-}\!\!>^2}$ are found to be
$$<\!\!n_{-}\!\!> = 33.84\pm0.23                  \eqno(7)$$
$$D_{-} = 26.36\pm0.68                    \eqno(8)$$
and their ratio is
$${{<\!\!n_{-}\!\!>}\over{D_{-}}} = 1.284\pm0.034  \eqno(9)$$
\smallskip
Figure 5 shows the multiplicity distribution for full phase space; again
the error bars are statistical.
The mean multiplicity, dispersion and the ratio are now
$$<\!\!n_{-}\!\!> \,\, = 57.01\pm0.39                            \eqno(10)$$
$$D_{-} = 44.04\pm0.82                                           \eqno(11)$$
$${{<\!\!n_{-}\!\!>}\over{D_{-}}} = 1.295\pm0.020                \eqno(12)$$
\smallskip
As this distribution is inferred from the one measured in the TPC, it may
depend on the model used for simulation [10]. A Monte-Carlo study of the
rapidity distribution suggests that the systematic error of $<\!\!n_-\!\!>$
due to extrapolation to full phase space is of the order of 3.5\%.

The corrected distributions have a maximum at low multiplicities corresponding
to peripheral collisions, followed by a somewhat flat region ('plateau')
for intermediate values of the impact parameter. At higher multiplicities a
steep fall off ('tail'), due to central collisions is observed.

The corrected distributions are of about the same shape as those measured
by other experiments, especially NA35 [16,17] and NA34 [18]. In particular, our 
distribution for S-Pb collisions is compatible with NA35 O-Au [16]
distribution when given in terms of KNO variables. One observes, however,
a discrepancy between the distributions in fig. 4 and fig. 5 and those
reported by WA80 [19]. This can be explained by the fact that peripheral
collisions were strongly suppressed in the data of ref. [19], whereas they
are included in the present analysis.

A proportionality between $D_{-}$ and $<\!\!n_{-}\!\!>$ (figure 6) from
collisions
of oxygen with different targets at energies of 60 and 200 $GeV/n$ were
reported in ref. [16]. Added in figure 6 are the results from S-S and
S-Cu collisions [17] as well as the present S-Pb measurement. All points
tend to lie on a straight line. A similar proportionality had been observed
long ago for pp collisions [20]. This proportionality was, in the case of
nucleus-nucleus collisions, explained in terms of a superposition of
independent nucleon-nucleus collisions [16].

Entropy is an alternative, useful variable for the study of multiparticle
production [21,22]; it reflects general features of independent particle
production. The total entropy from statistically independent phase space
regions, e.g. intervalls $\Delta y$, is given by the sum of the entropies
of these regions. Therefore, the total entropy $S$ is proportional to the
total rapidity range $Y_m$: $S \sim Y_m$. This is related to Feynman's argument 
on scaling in the variable $\xi = y/Y_m$ [21]. In addition, it follows
from eq. (4) that $S$ is invariant under distorsions of the multiplicity scale.
An analysis of hadron-hadron collisions for $\sqrt{s} > 20$ $GeV$ shows that
the entropy (4) increases linearly with the maximum rapidity $Y_m$ of the
produced pions [21]
$${S \over {Y_m}} = 0.417\pm0.009                 \eqno(13)$$
For S-Pb collisions in this experiment one finds
$${S \over {Y_m}} = 0.487\pm0.003                 \eqno(14)$$
where $\displaystyle Y_m = \ln\biggl({{\sqrt s - n_p m_n}\over m_{\pi}}\biggr)$, 
$m_{\pi}$ ($m_n$) is the pion (nucleon) mass and $\sqrt{s}\!$
\footnote*{$s = (\underline{P_A} + \underline{P_B})^2$, \hfill\break
where ${\underline{P_A}} = (\sqrt{A^2 P_L^2 + (A m_n + \epsilon_A)^2},
\overrightarrow{P_T} = \overrightarrow{0}, A \overrightarrow{P_L})$
and ${\underline{P_B}} = (\overline{B} {m_n} + \epsilon_B,
\overrightarrow{P} = \overrightarrow{0})$; \hfill\break
$\overline{B} = B - A \lbrack (B/A)^{2/3} - 1 \rbrack^{3/2}$ is the number of
participants in the target for central collisions. $A$ and $B$ are,
respectively, the numbers of nucleons of projectile (sulphur) and target
(lead), $P_L$ = 200 $GeV/c$ and $m_n$ is the rest mass of a nucleon; the
binding energies $\epsilon_A$ and $\epsilon_B$ were neglected.}
is the center of mass energy of $n_p\, (= A + \overline{B})$ participating
nucleons. The maximum rapidity is obtained for central collisions, therefore
$Y_m^{S-Pb}$ = 8.76 was used. (Note that $Y_m^{p-p}$ = 4.93 at 200 $GeV/c$).

\v{S}im\'{a}k et al. [23] show also that multiparticle production exhibits a
multifractal behaviour by investigating higher generalized fractal dimensions
of order $q$ given by
$$D_q = {I_q \over Y_m}                               \eqno(15)$$
where
$$I_q = {1 \over {1-q}}\,\ln \Bigl(\sum_n t_n^q\Bigr)       \eqno(16)$$ 
is the R\'enyi generalized entropy [24].
It can be shown that 
$$\lim_{q \to 1}{I_q} = S                     \eqno(17)$$
where $S$ is given by (4).
Figure 7 shows measured values of $D_q$ for different values of $q$. $D_q$ 
decreases with $q$, which may be considered as a signal of a multifractal 
behaviour in multiplicity distributions. This behaviour has also been found 
in ref. [25] for ion-emulsion interactions at various energies. The absolute 
values of $D_q$ measured are smaller than those of ref. [25], which may be due 
to the values of $Y_m$ for the AgBr target.

Generalized fractal dimensions $D_q$ were also determined from the newly 
introduced Tsallis generalized entropy [26]:
$$I_q = k\, {{1 - \sum_n t_n^q} \over {q - 1}}       \eqno(18)$$
with $k = 1$ [26]; they are  given in fig. 7 as well. (Note that $I_q$ in (18)
fulfills eq. (17)).
\bigskip\noindent
{\bf Conclusion.} 
\bigskip
A high statistics study of fully inclusive negative multiplicity distributions  
from S-Pb collisions at 200 $GeV/c$ is presented both for limited and full 
phase space.
A proportionality between $<\!\!n_{-}\!\!>$ and $D_{-}$ is observed. 
The generalized fractal dimensions are shown to decrease with 
increasing order which may be interpreted as a multifractal behaviour of 
the multiplicity distribution. 
\bigskip\noindent
{\bf Acknowledgements.}
\bigskip
One of us (M.H.) gratefully acknowledges helpful discussions with C. Fuglesang 
and M. Schmelling, and kind hospitality at IReS. Part of this work was 
supported by Director, Office of Energy Research, Division of Nuclear Physics 
Of the Office of High Energy and Nuclear Physics of the U. S. Department of 
Energy under contract no DE-AC03-76SF00098 (LBL), DE-FG02-91ER40652 
(Creighton), DE-FG02-87ER40315 (CMU), by the United Kingdom Science and 
Engineering Council under grant GR/F 40065 and by the EC under contract 
A88000145. Authors are grateful to the CERN Directorate, the former CERN EP, 
EF, DD and SPS divisions and the CERN's PPCS group, and especially to D. Lord.
\bigskip\noindent
{\bf Note added in proof.}
\bigskip
The multiplicity distribution presented here is well fitted by the percolating
string fusion model and by the generalized negative binomial model (S. Hegyi,
private communication). 
\vfill\eject
\noindent
{\bf References.}
\bigskip
\item{[1]} M. Hafidouni, Doctorat de l'Universit\'e Louis Pasteur de Strasbourg,
CRN/HE 92-35 (1992).
\item{[2]} C. Garabatos, Nucl. Instr. Meth. A283 (1989) 553.
\item{[3]} M. Aguilar-Benitez et al., Nucl. Instr. Meth. 258 (1987) 26.
\item{[4]} R.K. Bock, H. Grote, D. Notz and M. Regler, in: M. Regler (Ed.),
Data Analysis Techniques in High Energy Physics, Cambridge University Press, 
Cambridge, 1990.
\item{[5]} NA36 Collab., E. Andersen et al., Nucl. Instr. Meth. A301 (1991) 69.
\item{[6]} NA36 Collab., E. Andersen et al., Phys. Lett. B220 (1989) 328.
\item{[7]} C. Brechtmann and W. Heinrich, Z. Phys. A331 (1988) 463. 
\item{[8]} UA5 Collab., R.E. Ansorge et al., Z. Phys. C43 (1989) 357.  
\item{[9]} ALEPH Collab., D. Decamp et al., Phys. Lett. B273 (1991) 181. 
\item{[10]} J.-P. Pansart, Nucl. Phys. A261 (1987) 521c; Saclay Preprint DPhPE 
89-04 (March 1989); in: J. Tran Thanh Van (Ed.), Proceedings of the 23rd 
Rencontre de Moriond, Editions Fronti\`eres, Gif-sur-Yvette, 1988.  
\item{[11]} C.E. Shannon, Bell Syst. Tech. J. 27 (1948) 379. This paper 
is reproduced in: D. Slepian (Ed.), Key Papers in The Development of 
Information Theory, IEEE Press, New-York, 1974. 
\item{    } C.E. Shannon and W. Weaver, The Mathematical Theory of 
Communication, University of Illinois Press, Urbana, 1949.
\item{[12]} E.T. Jaynes, Phys. Rev. 106 (1957) 620; 108 (1957) 171. 
\item{[13]} J.E. Shore and R.W. Johnson, IEEE Trans. Inf. Th. IT-26 (1980) 26;
IT-29 (1983) 942.
\item{    } Y. Tikochinsky, N.Z. Tishby and R.D. Levine, Phys. Rev. Lett. 52 
(1984) 1357.
\item{[14]} C. Fuglesang, Nucl. Instr. Meth. A278 (1989) 765.
\item{[15]} C.S. Lindsey, Nucl. Phys. A544 (1992) 343c.
\item{[16]} NA35 Collab., A. Bamberger et al., Phys. Lett. B205 (1988) 583.
\item{[17]} NA35 Collab., J. B\"{a}chler et al., Z. Phys. C51 (1991) 157.
\item{[18]} A. Marzari-Chiesa, Nucl. Phys. A519 (1990) 435c.
\item{[19]} WA80 Collab., R. Albrecht et al., Z. Phys. C55 (1992) 539.
\item{[20]} A. Wr\'oblewski, Acta Phys. Pol. B4 (1973) 857. 
\item{[21]} V. \v{S}im\'{a}k, M. \v{S}umbera and I. Zborovsk\'{y}, 
Phys. Lett. B206 (1988) 159; in: O. Botner (Ed.), Proceedings of The 
International Europhysics Conference on High Energy Physics, European Physical 
Society, Petit-Lancy, Switzerland, 1987.
\item{[22]} P.A. Carruthers, M. Pl\"umer, S. Raha and R.M. Weiner, 
Phys. Lett. B212 (1988) 369.
\item{    } V. Majern\'{\i}k and B. Mamojka, Phys. Scripta 44 (1991) 412.
\item{[23]} M. Pachr, V. \v{S}im\'{a}k, M. \v{S}umbera and I. Zborovsk\'{y},
Mod. Phys. Lett. A7 (1992) 2333.  
\item{[24]} A. R\'enyi, Probability Theory, North-Holland, Amsterdam, 1970.
\item{[25]} A. Mukhopadhyay, P.L. Jain and G. Singh, Phys. Rev. C47 (1993) 410.
\item{[26]} Z. Dar\'oczy, Inf. Control 16 (1970) 36.
\item{    } C. Tsallis, J. Stat. Phys. 52 (1988) 479.
\item{    } E.M.F. Curado and C. Tsallis, J. Phys. A24 (1991) L69.
\vfill\eject 
\epsfbox{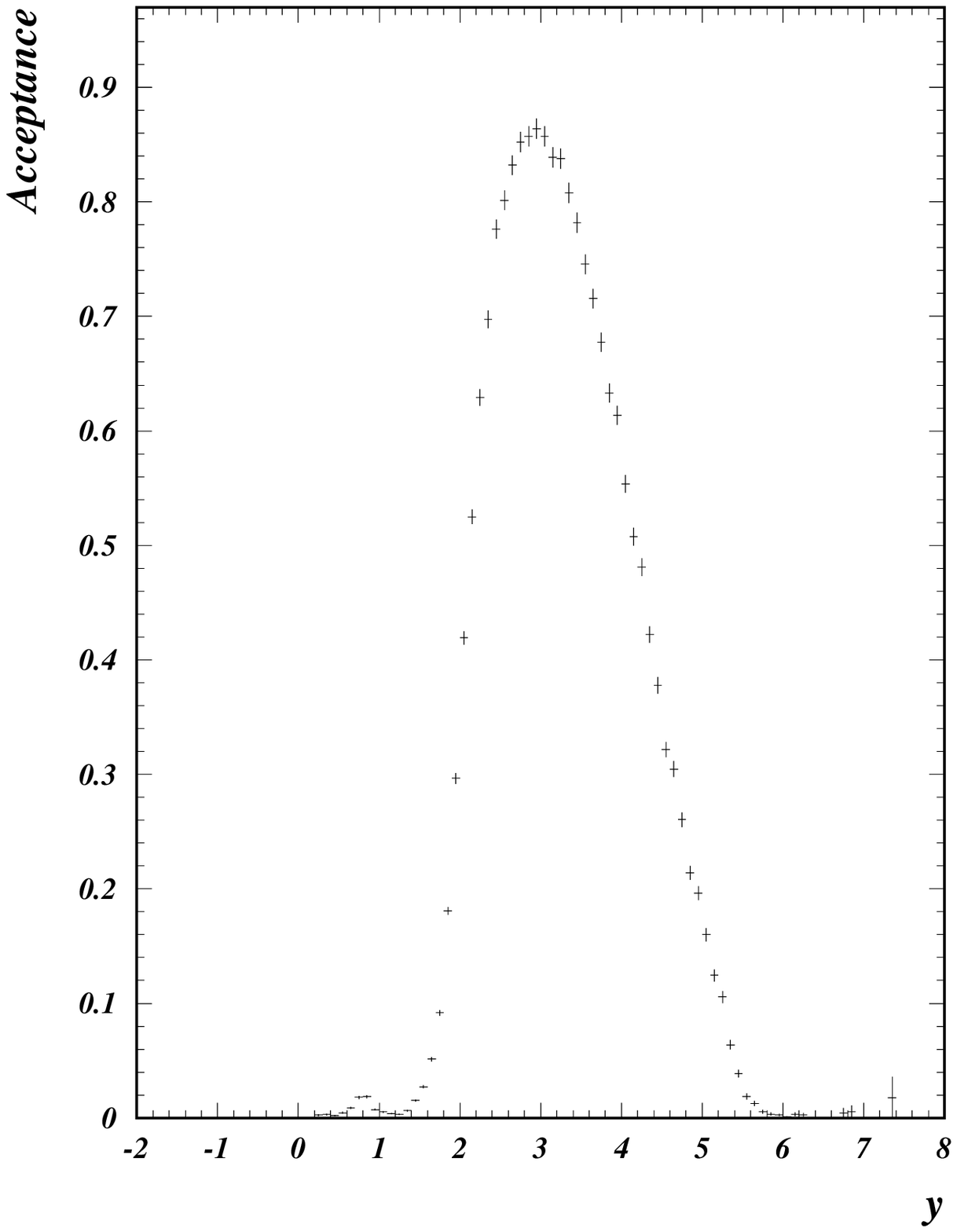}

\centerline{Fig. 1a. The acceptance of the TPC as a function of the laboratory
rapidity y}
\vfill\eject
\epsfbox{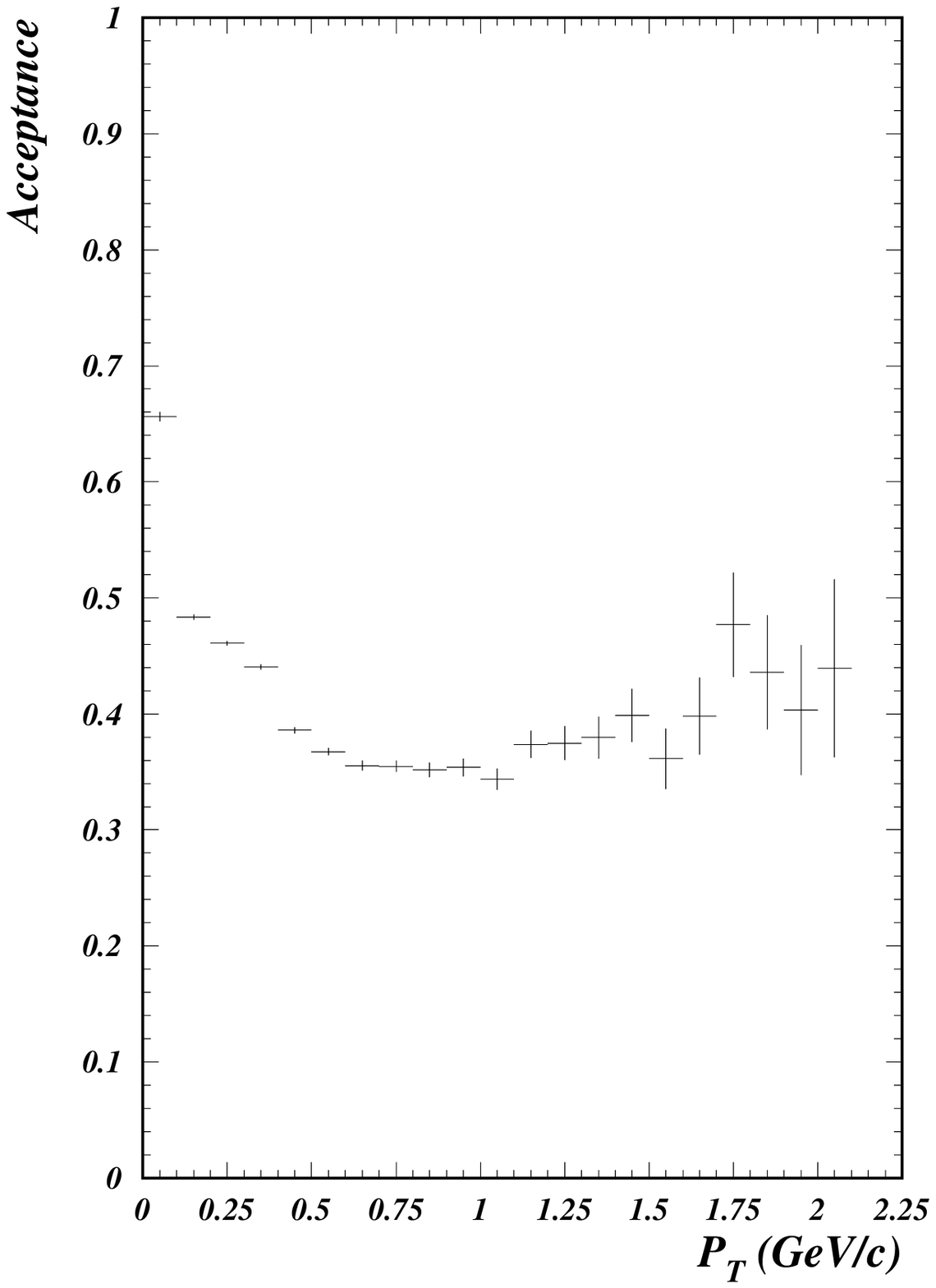}

\centerline{Fig. 1b. The acceptance of the TPC as a function of $P_T$}
\vfill\eject
\epsfbox{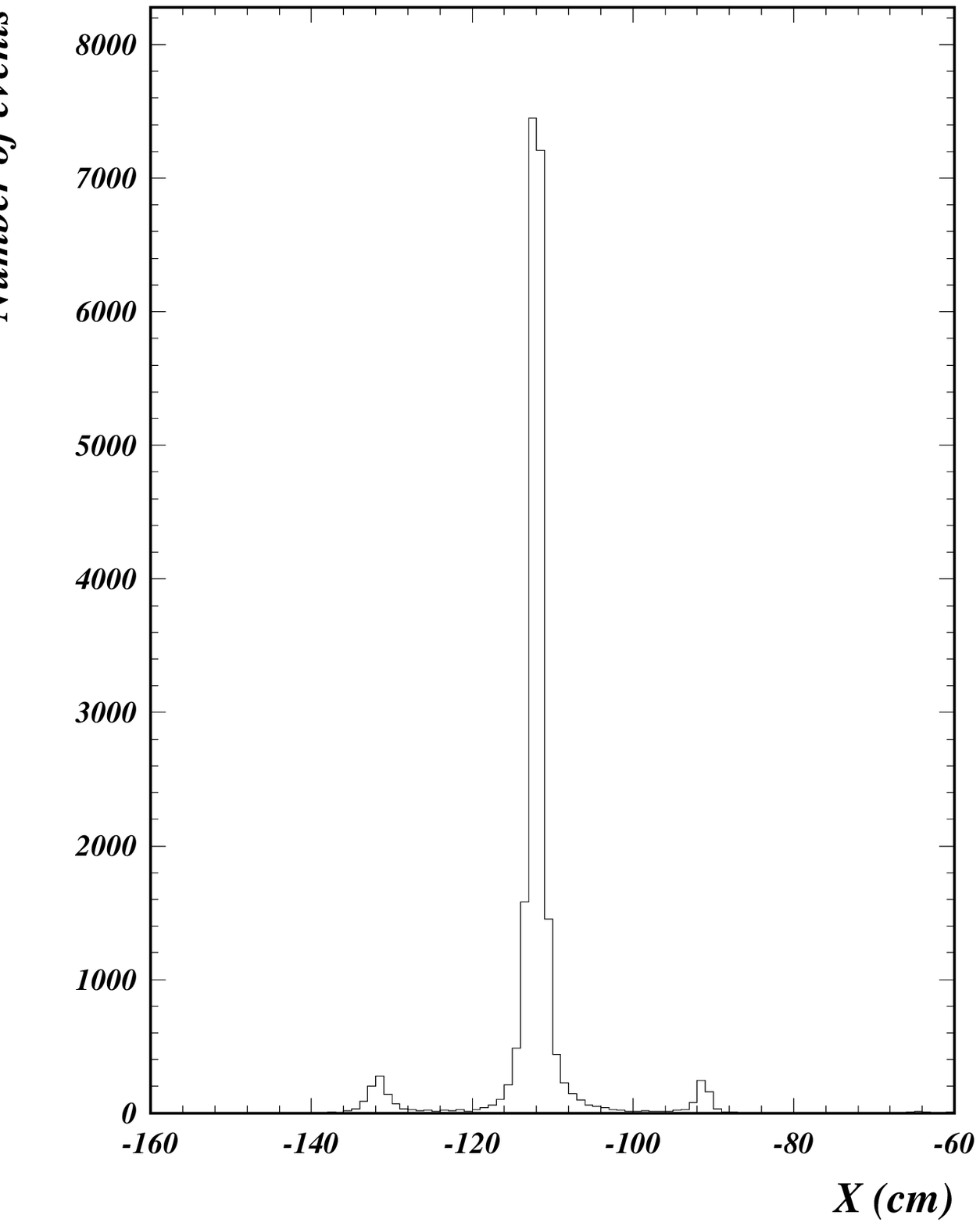}

\centerline{Fig. 2. Distribution of the vertex position X along the beam 
direction}
\vfill\eject
\epsfbox{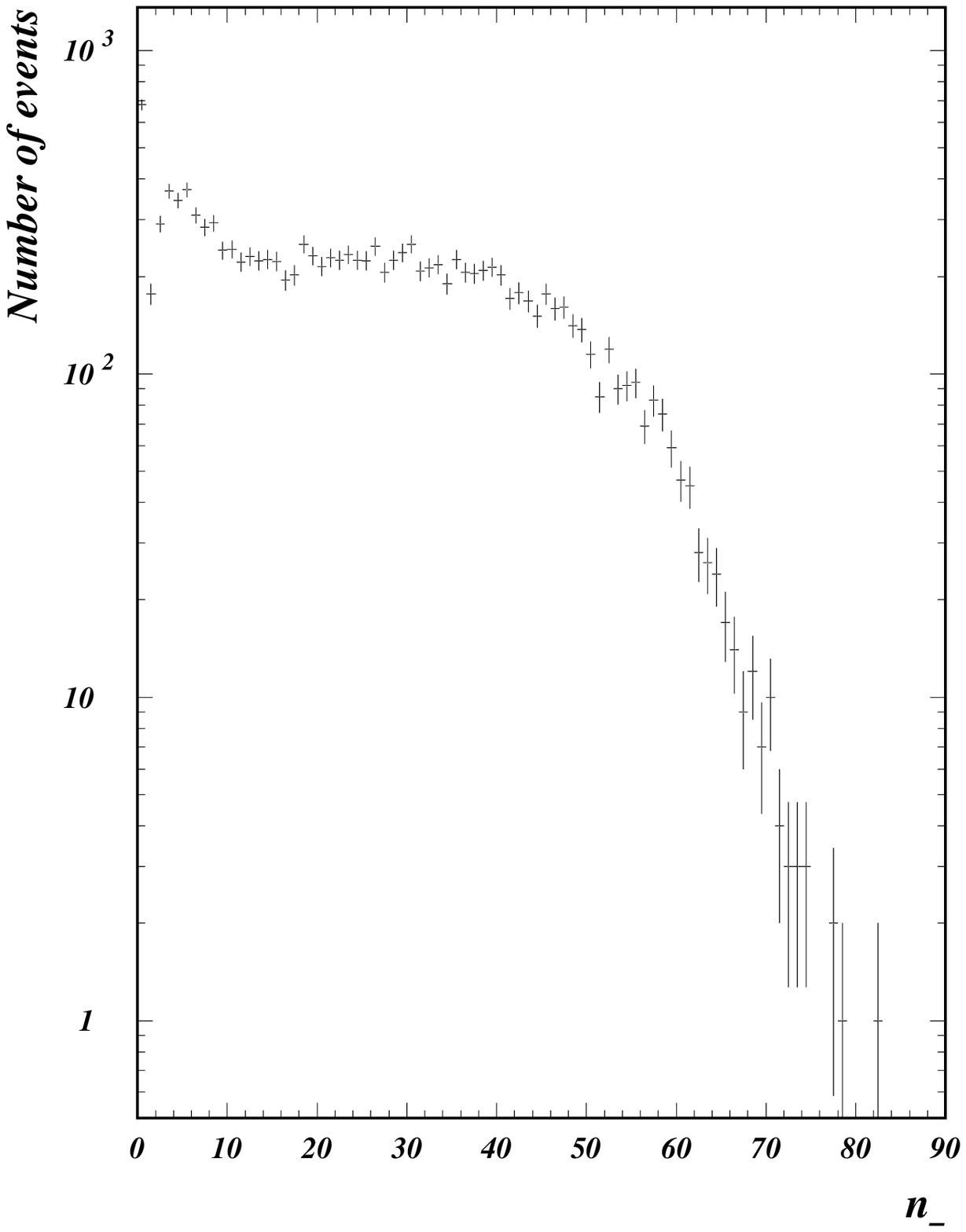}

\centerline{Fig. 3. Observed negative multiplicity distribution}
\vfill\eject
\epsfbox{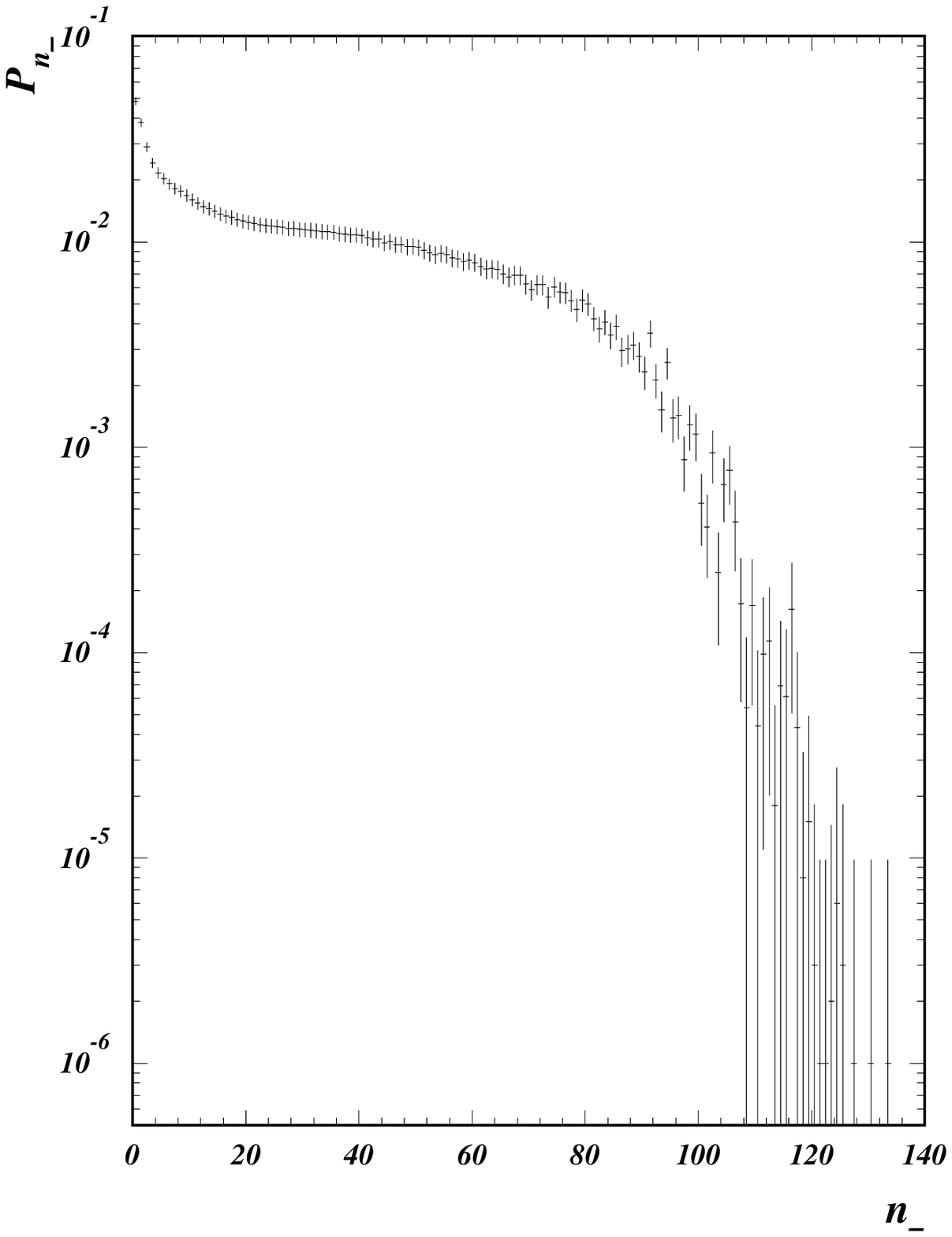}

\centerline{Fig. 4. The corrected negative multiplicity distribution 
for the TPC phase space}
\vfill\eject
\epsfbox{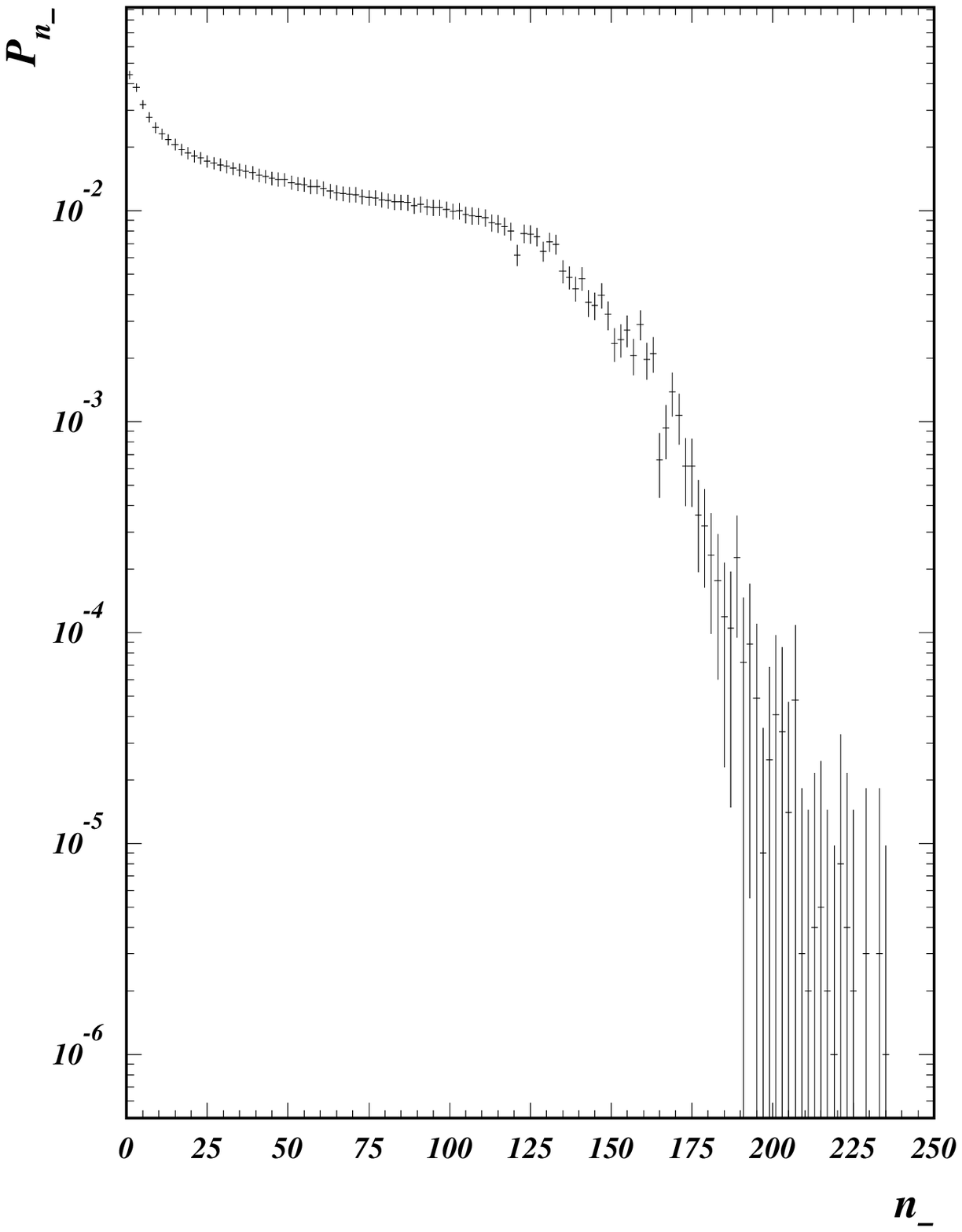}

\centerline{Fig. 5. The corrected negative multiplicity distribution for 
full phase space}
\vfill\eject
\hoffset=-1.5truecm
\epsfbox{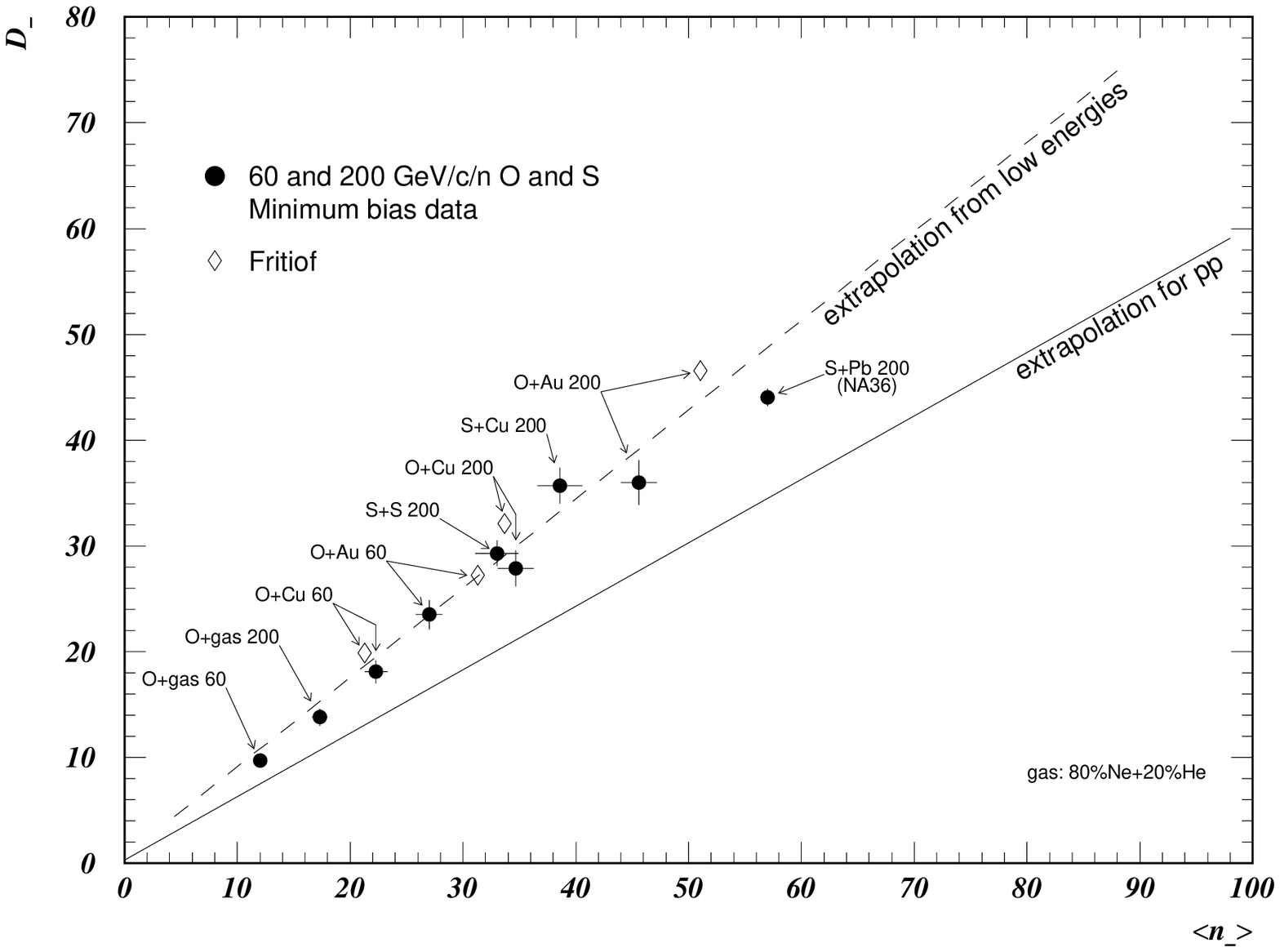}

\centerline{Fig. 6. $D_{-}$ as a function of $<\!\!n_{-}\!\!>$ for O and S 
projectiles and different targets,} 
\centerline{including the result of the present analysis of S-Pb interactions} 
\vfill\eject
\hoffset=1.5truecm
\epsfbox{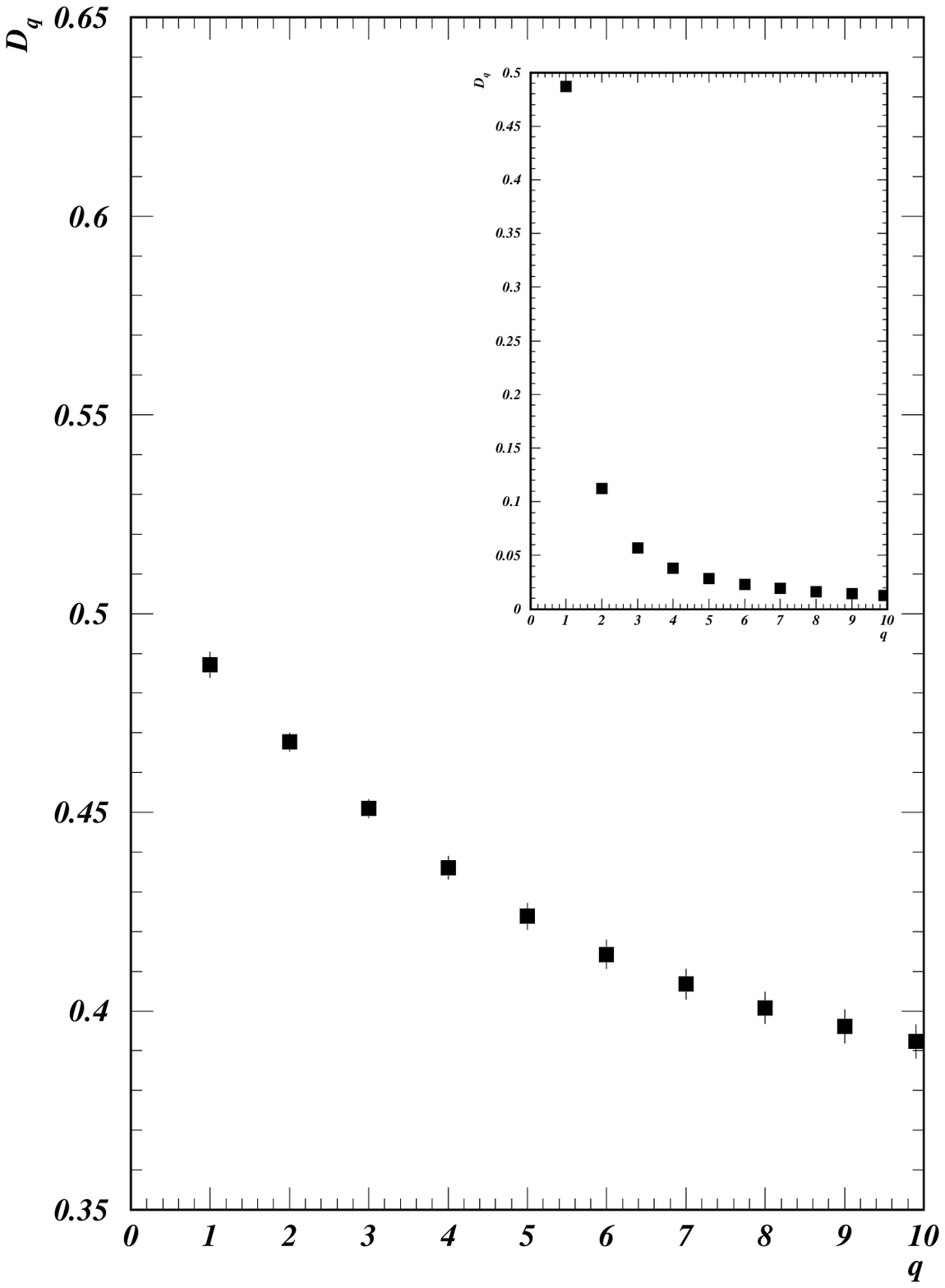}

\centerline{Fig. 7. Generalized fractal dimensions based upon Renyi generalized
entropy;}
\centerline{the insert gives those from Tsallis generalized entropy}
\bye